# Plasma density distribution and its perturbation by probes in axially symmetrical plasma

Valery Godyak[1] and Natalia Sternberg[2]

[1] RF Plasma Consulting, Brookline, Massachusetts 02446, USA. egodyak@comcast.net

[2] Department of Mathematics, Clark University, Worcester, Massachusetts 01610, USA

An analysis of plasma density distributions at arbitrary ion-atom collisionality for one-dimensional axially symmetrical cylindrical and annular plasmas is presented. Perturbations of plasma densities caused by a cylindrical probe are studied for arbitrary ion-atom collisionality. Analytical expressions for the plasma characteristics near the probe for low collisionality have been obtained. The plasma was modeled by the hydrodynamic neutral plasma equations, taking into account ionization, ion inertia, and a nonlinear ion frictional force, which dominates the plasma transport at low gas pressures. Significant plasma density depletion around the probe has been observed for a wide range of ion-atom collisionality. The presented results predict underestimation of plasma density obtained from the classical Langmuir probe procedure, and should provide a better understanding of electrostatic, magnetic, and microwave probes inserted into plasmas at low gas pressure.

## 1. Introduction

Plasma distribution in a gas discharge plasma is controlled by ionization, electron equilibrium with an ambipolar field, ion acceleration with the ambipolar field, and ion-neutral collisions. In the extreme case of no ion collisions, the plasma density distribution in planar and cylindrical geometry was given by the Tonks-Langmuir theory[1] for a bound gas discharge plasma (including a neutral plasma and sheaths on its boundaries). Later, Self [2,3] generalized the Tonks-Langmuir theory and presented a detailed analysis of gas discharge plasmas for different geometries and ionization mechanisms (uniform, direct, and two-step ionization).

For collision-dominated plasmas, the spatial distribution of the plasma characteristics was given by Schottky[4]. However, due to the zero-boundary condition for the plasma density, the Schottky model yields a singularity at the plasma boundary in the plasma velocity and in the plasma potential. All this makes the Schottky plasma model impossible to reconcile with a sheath model and to use in problems, which account for the interaction of electromagnetic fields with bounded plasmas. This was mitigated by Persson[5] who showed that in planar geometry, by taking into account ion inertia and collisions, one obtains at the plasma boundary a non-zero plasma density and a finite ion velocity equal to the ion sound speed, $v_s$.

Planar and cylindrical plasma models were also considered by Self and Ewald[6] for arbitrary collisionality and constant ion mobility (linear frictional force). They obtained an analytical solution for planar plasmas and a numerical solution for cylindrical plasmas from collisionless up to highly-collisional ion motions. In the limiting case of a non-collisional plasma, the plasma density profile is close to the classical Tonks-Langmuir distribution[1], while for a collision-dominated plasma, it is close to the Schottky diffusion distribution.[4]

The assumption of the constant ion mobility, used in Refs. 5,6, is applicable only at relatively high gas pressures when the ion drift velocity in the ambipolar field v is lower than the ion thermal velocity $v_{Ti}$. At lower gas pressures, the ion drift velocity exceeds the ion thermal velocity and reaches the ion sound speed $v_s \gg v_{Ti}$ at the plasma boundary. Thus, for the largest part of the plasma volume, $v > v_{Ti}$ holds, and the ion charge exchange governs the ion collision process. At such conditions, known as the regime of the variable ion mobility, the ion friction force is given by $F(v) = (\pi/2) Mv^2/\lambda_i$, where M is the ion mass and $\lambda_i$ is the ion mean free path, which is practically independent of the ion velocity. According to Refs. 7, 8, the charge exchange dominates in the ion transport when $\lambda_i/\Lambda > T_g/T_e$, where $T_g$ and $T_e$ are the gas and electron temperatures, and $\Lambda$ is the plasma characteristic size.

Analytical solutions of the planar and cylindrical plasma models in the regime of variable ion mobility (non-linear diffusion) extrapolated to the collisionless limit were given in Ref. 8 (see also Ref. 9). Simple analytical expressions for the ionization frequency and the relative plasma density at the plasma boundary were derived for planar and cylindrical plasmas as functions of the ion collisionality. The analytical solution[10,11] for the fluid planar plasma model, which accounts for ionization, ion inertia, and non-linear ion friction force, yields the values for the ionization frequency and the relative plasma density at the plasma boundary which practically coincides with those in Ref. 8.

Taking ion inertia and a non-linear ion friction force into account is essential in the evaluation of plasma perturbations caused by different kinds of plasma diagnostics probes. It is usually assumed that the plasma local parameters inferred from the probe measurements are not distorted by the presence of a probe. Such an assumption, however, may lead to erroneous results. In reality, when a probe is inserted into a plasma, it inevitably leads to perturbations of the plasma parameters around the probe. The effects of plasma perturbations by a probe are well known for the classical Langmuir probe, however, they are less understood for relatively bulky electrostatic analyzers, magnetic (B-dot), and microwave probes. [12,13]

Plasma perturbations caused by the immersion of Langmuir probes have been studied theoretically and experimentally for highly collisional (few Torr) plasmas in the regime of ambipolar diffusion with constant ion mobility.[14-16] These studies have shown a considerable drop in plasma density and electron temperature around the probe.

Plasma density distribution in a spherical plasma and plasma density depletion around a spherical probe immersed into a plasma have been obtained by solving the fluid equations, where ion inertia and a nonlinear ion friction force were taken into account.[17] It has been demonstrated there that a significant plasma density depletion up to a distance essentially exceeding the probe radius occurs.

As far as we know, the effects of plasma depletion around cylindrical probes commonly used in practice have not been studied for low-collisional plasmas. Neither have been studied plasma profiles for cylindrical and annular plasmas at low collisionality. This is, however, exactly the regime where electrostatic probes, B-dot probes, and different kinds of microwave probes have been used, and is the topic of the present paper. We have solved the fluid axially symmetric neutral plasma model with ion inertia and variable ion mobility, in order to study the plasma perturbation caused by a cylindrical probe. We have also studied plasma density distributions in cylindrical and annular plasmas for arbitrary collisionality.

The paper is organized as follows. In Section II, we analyze a one-dimensional cylindrical neutral plasma model with ion inertia and a non-linear ion friction force. In Section III, we present a similar analysis for an annular plasma. In Section IV, we consider plasma perturbation by a cylindrical probe and conclude the paper with a discussion.

## II.   Fluid model of 1-D cylindrical plasma at variable ion mobility

Consider a bounded cylindrical neutral plasma ($n_e = n_i = n$) which satisfies the following conditions: a Maxwellian electron energy distribution, cold ions ($T_i << T_e$), no heat transfer ($\nabla T_e = 0$), and no net current along the plasma density gradient. We assume axial symmetry with respect to the axis which passes through the plasma center. We further assume that variations in the axial direction can be neglected. Moreover, we assume that the ions are accelerated by the ambipolar electric field towards the wall, while the electrons are in Boltzmann equilibrium with the radial ambipolar field.

Such plasma can be described quite accurately by the following system of hydrodynamic equations[5,6], namely by the continuity and momentum equations for ions, and the Boltzmann equilibrium for electrons in the ambipolar field:

$$\frac{1}{r}\frac{d(rnv_r)}{dr} = Zn \qquad (1)$$

$$Mv_r\frac{dv_r}{dr} + MZv_r + e\frac{dV}{dr} + \frac{\pi}{2\lambda_i}M|v_r|v_r = 0 \qquad (2)$$

$$kT_e\frac{dn}{dr} = en\frac{dV}{dr} \qquad (3)$$

where $r \geq 0$ is the radial distance, $n$ is the plasma density, $v_r$ is the plasma transport velocity, $V$ is the electric potential, $M$ is the ion mass, $T_e$ is the electron temperature, $e$ is the electron charge, $k$ is the Boltzmann constant, $Z$ is the frequency of direct ionization. The last term in eq. 2 is the non-linear ion friction force, $F_i \propto v_r^2$.

Hence, we can prescribe the following initial conditions at the plasma center:

$$n(0) = n_0 \quad v_r(0) = 0 \quad V(0) = 0 \qquad (4)$$

To simplify the computations, we introduce the following dimensionless variables:

$$x = \frac{r}{R} \quad y(x) = \frac{n(r)}{n_0} \quad u(x) = \frac{v_r(r)}{v_s} \quad \eta(x) = -\frac{e}{kT_e}V(r) \quad (5)$$

where R is the plasma radius and $v_s = (kT_e/M)^{1/2}$ is the ion sound speed. In those variables, system (1–3) becomes:

$$\frac{1}{x}\frac{d(xyu)}{dx} = Sy \qquad (6)$$

$$u\frac{du}{dx} + Su - \frac{d\eta}{dx} + \frac{\pi}{2}\alpha|u|u = 0 \qquad (7)$$

$$\frac{dy}{dx} = -y\frac{d\eta}{dx} \qquad (8)$$

where $0 \leq x \leq 1$, $S = ZR/v_s$ is the normalized frequency of ionization and $\alpha = R/\lambda_i$ is the collision parameter. From (4), we find the initial conditions at the plasma center $x = 0$:

$$y(0) = 1 \quad u(0) = 0 \quad \eta(0) = 0 \qquad (9)$$

In explicit form, equations (6-8) become

$$\frac{d\eta}{dx} = \frac{1}{1-u^2}\left(2Su - \frac{u^2}{x} + \frac{\pi}{2}\alpha|u|u\right) \qquad (10)$$

$$\frac{du}{dx} = \frac{1}{1-u^2}\left(S(1+u^2) - \frac{u}{x} + \frac{\pi}{2}\alpha|u|u^2\right) \qquad (11)$$

$$y = \exp(-\eta) \qquad (12)$$

First note that equations (10–12) have a singularity at $x = 0$. In our computations, we therefore use initial conditions at $x = \delta \approx 0$

$$\eta(\delta) = \frac{3}{8}S^2\delta^2 \qquad u(\delta) = \frac{S}{2}\delta \qquad (13)$$

which are obtained by differentiating (6) and (7) and then using Taylor expansions.

Furthermore, equations (10–11) also have a singularity at $u = 1$ (i.e., $v_r = v_s$). Because of that singularity, it is assumed that the plasma boundary $r = R$ (or $x = 1$) is located at the point where the ion velocity reaches the ion sound speed. This singularity, however, can be removed by considering $u$ as the independent variable and solving the equivalent system

$$\frac{d\eta}{du} = \frac{2Su - u^2/x + 0.5\pi\alpha|u|u}{S(1+u^2) - u/x + 0.5\pi\alpha|u|u^2} \qquad (14)$$

$$\frac{dx}{du} = \frac{1-u^2}{S(1+u^2) - u/x + 0.5\pi\alpha|u|u^2} \qquad (15)$$

$$y = \exp(-\eta) \qquad (16)$$

With the following initial condition at $u = S\delta/2$:

$$\eta\left(\frac{S}{2}\delta\right) = \frac{3}{8}S^2\delta^2 \qquad x\left(\frac{S}{2}\delta\right) = \delta \qquad (17)$$

We have solved the system (14–16) with the initial conditions (17) for different values of the collision parameter $\alpha$. The ionization frequency $S$ was determined through iterations using the condition at the plasma boundary where $u = 1$ and $x = 1$. The obtained normalized radial plasma density distributions $y = n/n_0$ as functions of $x$ for the collision parameters $\alpha = R/\lambda_i = 0$; $1$; $10$; and $100$ are shown in Fig.1. In general, the plasma densities decrease as $x$ increases. With growing $\alpha$, the plasma profiles become steeper near the plasma boundary.

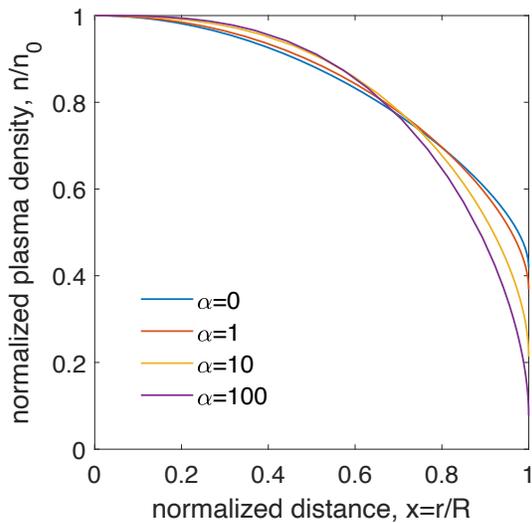

Fig. 1. Normalized radial plasma profile $y(x) = n/n_0$ for different collisional parameter $\alpha = R/\lambda_i$.

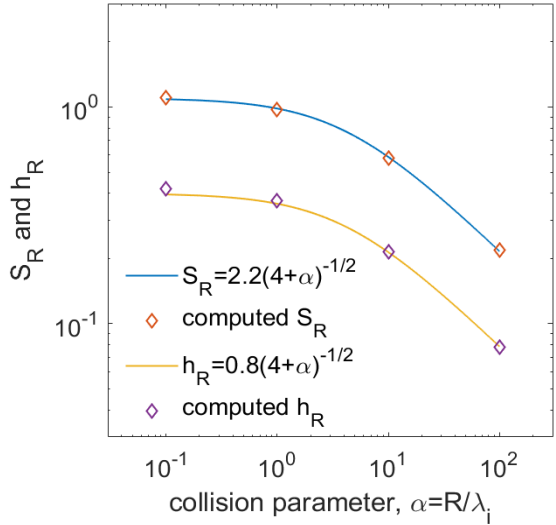

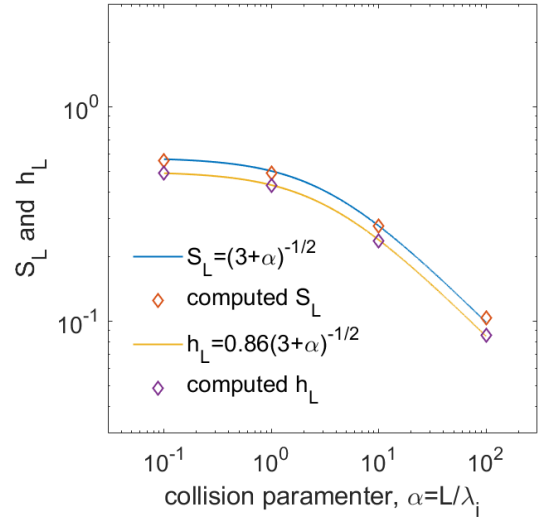

Fig. 2. Normalized ionization frequency, $S_R$ and normalized plasma density at the plasma boundary, $h_R$ in cylindrical plasma.

Fig. 3. Normalized ionization frequency, $S_L$ and normalized plasma density at the plasma boundary, $h_L$, in planar plasma.

The relative plasma densities at the plasma boundary $n(R)/n_0 = h_R$ and the normalized ionization frequency $S_R = ZR/v_s$ are shown in Fig. 2. The curves are obtained from the simplified heuristic model[8], while the discrete points are obtained from our numerical solutions of system (14 – 16). For the planar case, the relative plasma density $n(L)/n_0 = h_L$ at the plasma boundary $L$ and the normalized frequency of ionization $S_L = ZL/v_s$ are shown in Fig. 3. As before, the curves correspond to the heuristic model[8], while the discrete points were obtained by solving numerically the planar plasma problem in the variable ion mobility regime.[10,11] As one can see, qualitatively, the plasma densities at the plasma boundary and the frequencies of ionization in cylindrical and planar geometries behave in a similar manner. In both geometries, for $\alpha \ll 1$, the values of $h$ and $S$ do not change much, and remain close to the corresponding values in the collisionless case ($\alpha = 0$). For $\alpha \gg 1$, the corresponding curves drop as $\alpha^{-1/2}$.

An analytical solution of the 1-D models for planar and cylindrical plasmas in the regime of variable ion mobility (non-linear diffusion) has been obtained in Refs. 8,9. The problem was formulated by the continuity, Boltzmann, and ion mobility equations:

$$\nabla(nv) = Zn \quad n/n_0 = \exp(eV/kTe) \quad v = (2\lambda_i eE/\pi M)^{1/2} \tag{18}$$

where V and $E = -\nabla V$ are the electrical potential and ambipolar electric field.

A generalization of this model to the collisionless limit was done by introducing some effective ion mean free path $\lambda^{-1} = \lambda_i^{-1} + \lambda_s^{-1}$, where $\lambda_s$ corresponds to the ionization frequency of the Tonks-Langmuir model[1]. Solutions, satisfying $n(0)/n_0 = 1$ at the plasma center and $v = v_s$ at the plasma boundary, were obtained for the 1-D planar and cylindrical cases with the following analytical expressions for the normalized ionization frequency and the normalized plasma density at the plasma boundary:[8,9]

$$S_L = (3 + L/\lambda_i)^{-1/2} \quad \text{and} \quad h_L = 0.86\ (3 + L/\lambda_i)^{-1/2} \qquad \text{with } S_L/h_L = 1.16 \qquad (19)$$

for the planar plasma, and

$$S_R = 2.2\ (4 + R/\lambda_i)^{-1/2} \quad \text{and} \quad h_R = 0.80\ (4 + R/\lambda_i)^{-1/2} \quad \text{with } S_R/h_R = 2.75 \qquad (20)$$

for the cylindrical plasma.

The comparison of the h and S values given in (19 – 20) for planar and cylindrical plasmas in the range $10^{-1} \leq \lambda_i/L,\ R \leq 10^2$ with the corresponding values found numerically has demonstrated excellent agreements, as illustrated in Figs. 2 and 3. Thus, the analytical expressions (19) and (20) are quite accurate descriptions of S and h given by the fluid models which account for ion inertia and a non-linear ion friction force.

## III.   Fluid model for annular cylindrical plasma at variable ion mobility

Consider an annular axially symmetrical plasma confined between two coaxial cylindrical surfaces. As in Section II, we assume that only variations in the radial direction need to be taken into account. In this case, the neutral plasma is described by the same equations (1–3) as in Section II. However, we have to consider two neutral plasma boundaries: the inner boundary at $r = \rho$ along the inner cylinder, and the outer boundary at $r = R$ along the wall of the outer cylinder, $\rho < R$.

At each neutral plasma boundary, the ions reach the ion sound speed, $v_s$. In particular, $v_r(\rho) = -v_s$, and $v_r(R) = v_s$. It holds at some point $r = r_0$, $r < r_0 < R$, that $v_r(r_0) = 0$, $V(r_0) = 0$, and the plasma density achieves its maximum $n(r_0) = n_{max}$. That point $r_0$ depends not only on $\rho$, but also on collisionality.

To compute the plasma characteristic, we use the normalization (5) with $n_0 = n_{max} = n(r_0)$. The corresponding values at the inner plasma boundary $x = \rho/R$ then become

$$u(\rho/R) = -1 \quad \eta(\rho/R) = \eta_\rho \quad y(\rho/R) = \exp(-\eta_\rho) \tag{21}$$

Since $\eta_\rho$ is not known, we set

$$\eta = \tilde{\eta} + \eta_\rho \tag{22}$$

and find $\tilde{\eta}$ by solving 14–16 for $-1 \leq u \leq 1$, with the initial condition

$$x(-1) = \rho/R \qquad \tilde{\eta}(-1) = 0 \tag{23}$$

Note that $\tilde{\eta} = -\eta_\rho$ when $\eta = 0$, which occurs at $x_0 = r_0/R$, where both $\eta$ and $\tilde{\eta}$ are at the minimum, $u(x_0) = 0$ is the inflection point, and $y(x_0) = 1$ is the maximum.

In our computations, we have considered $\rho/R$ and $\alpha$ as parameters, while the frequency of ionization $S$ has been found though iterations by using the boundary condition at the outer cylinder $x(1)=1$. Our results are presented in Figs. 4-7 where the normalized radial plasma density distributions in an annular plasma are shown for different ratios $\rho/R = 0.1;\ 0.2;\ 0.3;\ 0.5;\ 0.7;\ 0.9$ and different collision parameters $\alpha = 0;\ 0.1;\ 1.0;\ 10$ and $100$.

A common feature of these plasma profiles is the shift of the plasma density maximum to the position of the inner plasma boundary ($r = \rho$) as $\rho/R$ is decreasing. The plasma flattering at large collisionality is clearly seen in Figs. 6 and 7. Almost symmetrical distributions for large ratios $\rho/R = 0.7$ and $\rho/R = 0.9$ are observed, and in those cases, the plasma profiles are very close to the corresponding plasma profiles in a planar plasma with $L = 0.5(R-\rho)$.

The normalized plasma densities at the plasma boundaries, $n(R)/n_{max}$ and $n(\rho)/n_{max}$ are shown in Figs. 8 and 9 as functions of the ratio $\rho/R$. At the outer boundary, the values of $n(R)/n_{max}$ are monotone for all collision parameters $\alpha$, while the values at the inner boundary have minima for large collision parameters (see $\alpha = 10$ and $\alpha=100$). For all collision parameters, in the limit $\rho/R \to 1$, the plasma densities at each boundary converge to $h_L=0.5$, which is precisely the normalized plasma density at the plasma boundary for symmetrical planar plasmas[6, 10,11]. For example, for $\alpha=1$ and $\rho/R = 0.9$, we have $n(R)/n_{max}=0.491 \approx 0.5 \approx n(R)/n_{max}= 0.499$.

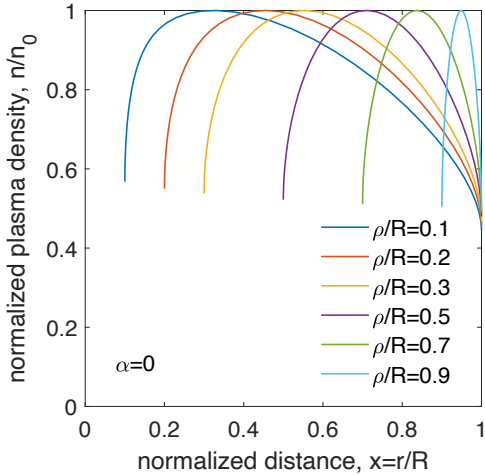

Fig. 4. Normalized radial plasma density profile, y = n/n₀ in collisionless annular plasma (a = 0).

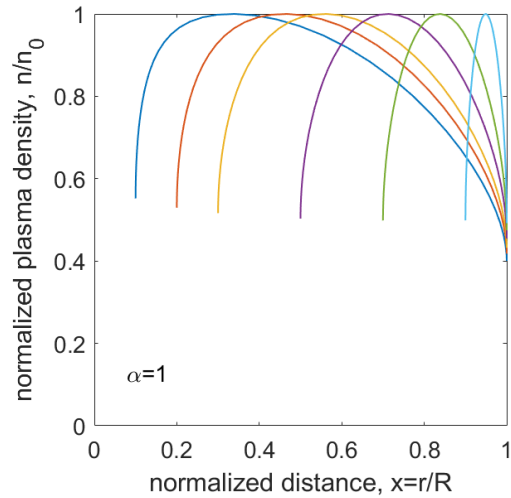

Fig. 5. Normalized radial plasma density profile, y = n/n₀ in annular plasma at $\alpha$ = 1.

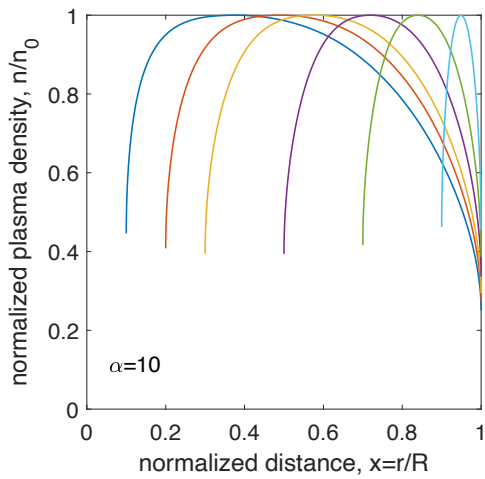

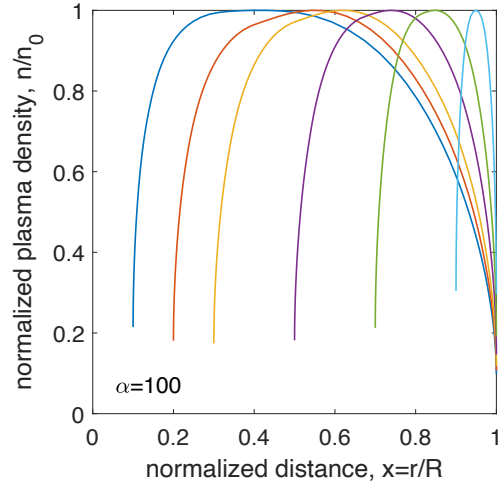

Fig. 6. Normalized radial plasma density profile, y = n/n$_0$ in annular plasma at $\alpha = 10$.

Fig. 7. Normalized radial plasma density profile, y = n/n$_0$ in annular plasma at $\alpha = 100$.

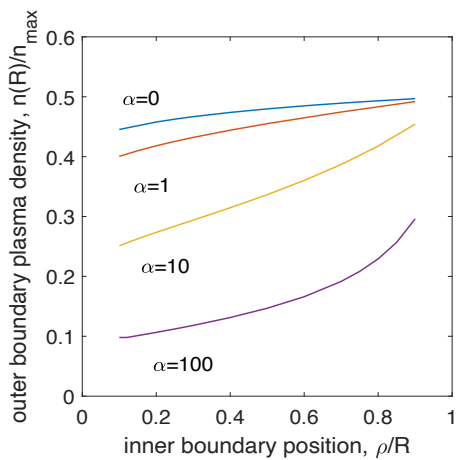

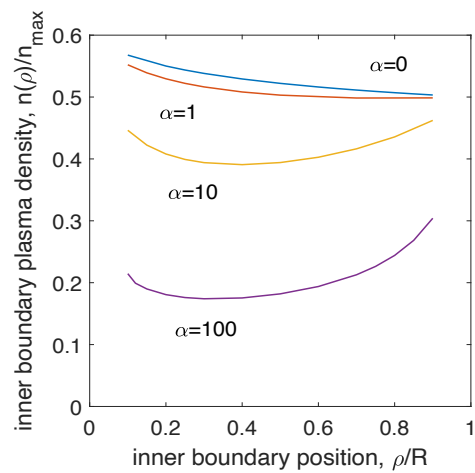

Fig. 8. Normalized plasma density at outer radial boundary at r = R as functions of $\rho/R$ at different collision parameter $\alpha$.

Fig. 9. Normalized plasma density at inner radial boundary r = $\rho$ as functions of $\rho/R$ at different collision parameter $\alpha$.

## IV. Plasma density depletion around a cylindrical probe

Many diagnostic tools for the experimental study of low-temperature plasmas use probes inserted into the plasma. When a probe is inserted into a plasma, it inevitably leads to perturbations of the plasma parameters around the probe, which may distortions of the probe measurements. Recently, concerns were raised about the validity of plasma parameters inferred from measurements by magnetic and microwave probes when plasma perturbations by the probe were not taken into account.[12,13]

For electrostatic, microwave, and magnetic probes, the main interaction of the electromagnetic field with the plasma takes place in the near-zone around the probe within a distance of a few probe radii. This is precisely the region where the plasma density is depleted by the probe. Indeed, at gas pressures typical in contemporary plasma applications and in basic plasma experiments, where ion inertia and ion charge exchange dominate the plasma transport, a spherical probe causes a significant plasma density depletion up to the distance essentially exceeding the probe radius.[17] Thus, the plasma density and the rf magnetic field inferred from the measurements with those probes correspond to a reduced plasma density and a distorted plasma rf current around magnetic probes. They do not provide accurate measurements for an unperturbed plasma density, or a plasma rf current in the absence of the probe.

In practice, Langmuir, magnetic, hairpin, and cutoff microwave probes are usually made of a thin wire with the ratio $a_p/l_p \approx 1x10^{-2} << 1$, where $a_p$ is the probe radius and $l_p$ is its length. This suggests that a 1-D model is quite adequate for analyzing the plasma depletion effect caused by a cylindrical probe. Plasma perturbations by a cylindrical probe can be modeled as a plasma nested between two cylinders, using the approach of Section III. However, in this case, since $\rho = a_p + s << R$, where $s$ is the width of the sheath along the probe, we need to "zoom in" on the area close to the probe. This is achieved by normalizing the distance to the inner plasma boundary $r = \rho$ and by changing the variables as follows:

$$\xi = \frac{r}{\rho} \quad y(\xi) = \frac{n(r)}{n_0} \quad u(\xi) = \frac{v_r(r)}{v_s} \quad \eta(\xi) = -\frac{e}{kT_e}V(r)$$

(24)

The system (14-15) then becomes

$$\frac{d\eta}{du} = \frac{2S_\rho u - u^2/\xi + 0.5\pi\beta|u|u}{S_\rho(1+u^2) - u/\xi + 0.5\pi\beta|u|u^2}$$

(25)

$$\frac{d\xi}{du} = \frac{1-u^2}{S_\rho(1+u^2) - u/\xi + 0.5\pi\beta|u|u^2}$$

(26)

where $S_p = Z\rho/v_s$ and b = $\rho/\lambda_i$. Note that at this normalization, $1 \leq \xi \leq R/\rho$, and at the plasma boundaries, $-u(1) = u(R/\rho) = 1$. Furthermore, at the point $\xi_0 = r_0/\rho$ where the plasma density is maximal:

$$u(\xi_0) = 0 \quad \eta(\xi_0) = 0 \quad y(\xi_0) = 1$$

(27)

For our computations, we have chosen $\rho/R = 10^{-3}$. Note that for such $\rho << R$, the plasma is unperturbed and uniform at some distance of several $\rho$ away from the probe.

For $\rho << R$ and $\beta << 1$ the system (25-26) can be reduced to:

$$\frac{d\eta}{du} = u \tag{28}$$

$$\frac{d\xi}{du} = -\frac{(1-u^2)\xi}{u} \tag{29}$$

which yields the analytical solution:

$$\xi = |u|^{-1}\exp\left(\frac{u^2-1}{2}\right) \tag{30}$$

$$\eta = \frac{1}{2}u^2 \tag{31}$$

$$y = \exp(-\eta) \tag{32}$$

Fig. 10 shows the computed plasma density distributions perturbed by a cylindrical probe, together with the analytical solution given by (30-32), and the normalized distribution of the microwave field $E/E_0$, generated by the probe. For comparison, Fig. 11 shows the computed plasma density distributions and the normalized microwave field for a spherical probe.[17] It is obvious from these figures that in both cases, the area of the microwave field concentration overlaps with the area of the maximal plasma density depletion. Qualitatively, plasma density perturbations for both probes are similar, but the plasma density depletion (caused by a spherical probe) spreads over a shorter distance from the inner plasma boundary than for a cylindrical probe, which is expected.

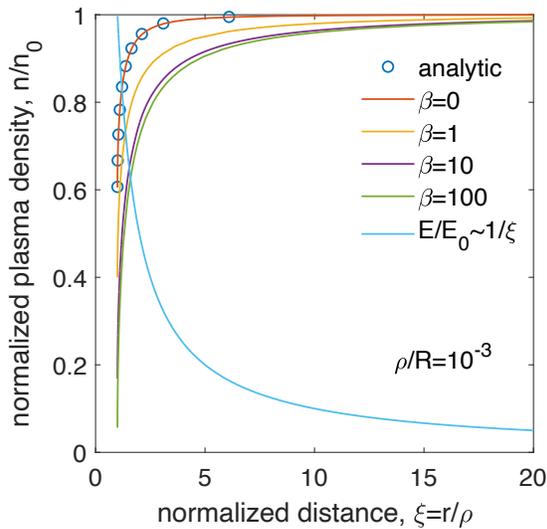

Fig. 10. Plasma density depletion around a cylindrical probe for different collision parameter $\beta = \rho/\lambda_i$ as functions of the normalized distance $\xi$.

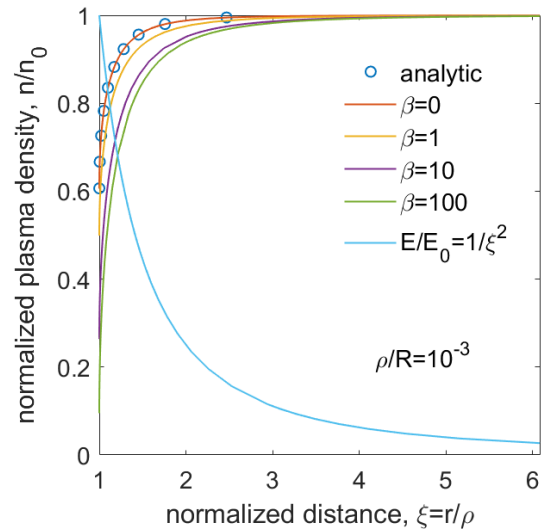

Fig. 11. Plasma density depletion around a spherical probe for different collision parameter $\beta = \rho/\lambda_i$ as functions of the normalized distance $\xi$.

In real probe experiments, $\rho/R << 1$, and $\beta = \rho/\lambda_i << \alpha = R/\lambda_i$. Therefore, a near collisionless regime in the plasma around a diagnostic probe is realized at moderate collisionality of the gas discharge plasma. This allows to account for plasma depletion around the probe using an analytical solution given for cylindrical probes by (30-32) for $\beta << 1$ when collisions can be neglected. The analytical solution for spherical probes at $\beta << 1$ is given in Ref. 17.

The values of the normalized plasma density at the plasma boundary along the probe ($\xi = 1$) are presented in Table 1 for cylindrical and spherical probes for different collision parameter $\beta$.

| $\beta = \rho/\lambda_i$ | 0 | 1 | 10 | 100 |
|---|---|---|---|---|
| Cylindrical $n_\rho/n_0$ | 0.605 | 0.400 | 0.169 | 0.0564 |
| Spherical $n_\rho/n_0$ | 0.606 | 0.499 | 0.264 | 0.0952 |

Table 1. Normalized plasma density at its boundary, n /n$_0$ for different collision parameters, $\beta$.

## V. Discussion

The plasma density profiles of the cylindrical and annular plasma have been obtained for a 1-D fluid model of low-pressure discharge plasmas in the regime of variable ion mobility. The plasma depletion caused by a cylindrical probe has been analyzed, using a 1-D fluid model, in the frame of the considered fluid model and an analytical solution for the low-collisional regime has been obtained. The numerical and analytical computations have shown that plasma density depletion always occurs around a probe and increases with ion-atom collisions.

Plasma depletion around the probe also affects the thickness of the sheath along the probe surface, which further affects the measurements with microwave probes. In addition, due to electron thermal motion near the probe surface, the gradient of the plasma density can affect the resonance frequency of the probe, and lead to additional dissipative effects.[12]

It is well-known that the validity of the Langmuir probe theory is limited by the condition $a_p << \lambda_e$, where $a_p$ is the probe radius and $\lambda_e$ is the electron mean free path. In the classical Langmuir probe, the electron depletion around the probe immersed into a uniform plasma is automatically accounted for in the collisionless probe theory when $a_p << \lambda_e$. Since in gas discharge plasmas $\lambda_i << \lambda_e$, the plasma density depletion may be essential, when $\lambda_i \leq \rho = a_p + s$, even for $\lambda_e >> a_p$. Here, $s$ is the width of the sheath along the probe.

For a typical plasma experiment in argon gas, $a_p = 0.1\ mm$, $kT_e = 3\ eV$, $n = 3 \times 10^{10}\ cm^{-3}$, and suppose $s = 3\lambda_D$, where $\lambda_D$ is the Debye length. Then for $\lambda_i \approx \rho$ (which corresponds to argon gas pressure of $p \approx 100\ mTorr$ with $\lambda_i \approx 3.3\mathrm{x}10^{-2}$ cm and $\lambda_e \approx 1\ cm >> a_p = 10^{-2}$ cm), the underestimation in the plasma density measured with the Langmuir probe is expected to be about 50%. The possibility of such a scenario deserves a thorough analysis at the kinetic level for both, electrons and ions. It may resolve the long-standing problem of why plasma densities obtained with Langmuir probes are consistently lower than those obtained in microwave diagnostics.

## Data availability

The data that support the findings of this study are available from the corresponding author upon reasonable request.